\documentclass{desyproc}
\usepackage{mciteplus}
\usepackage{xcolor}
\definecolor{lcolor}{rgb}{0.5,0,0}
\definecolor{citcolor}{rgb}{0,0.3,0.0}

\usepackage[breaklinks,colorlinks,urlcolor=blue,citecolor=citcolor,linkcolor=lcolor]
{hyperref}

\begin{document}

\newcommand{\sigmadip}{{ \sigma_{\textnormal{dip}} }}
\newcommand{\dsigmaa}{ { \frac{\ud \sigma^A_{\textnormal{dip}}}{\ud^2 \bt} } } 
\newcommand{\Aavg}[1]{\left\langle #1 \right\rangle_{\textnormal{N}}}
\newcommand{\A}{{\mathcal{A}}}
\newcommand{\xpom}{{x_{\mathbb{P}}}}
\newcommand{\dsigmap}{{\frac{\ud \sigma^{\textnormal{p}}_{\textnormal{dip}}}{\ud^2 \bt}}}
\newcommand{\ampli}{{\mathcal{N}}}
\newcommand{\sigmap}{{ \sigma^\textnormal{p}_{\textnormal{dip}} }}

\newcommand{\ud}{\, \mathrm{d}}

\newcommand{\ot}{{\mathbf{0}_T}}
\newcommand{\qt}{{\mathbf{q}_T}}
\newcommand{\rt}{{\mathbf{r}_T}}
\newcommand{\xt}{{\mathbf{x}_T}}
\newcommand{\bt}{{\mathbf{b}_T}}
\newcommand{\Deltat}{\boldsymbol{\Delta}_T}

\newcommand{\nc}{{N_\textnormal{c}}}
\newcommand{\qs}{Q_{\textnormal{s}}}
\newcommand{\as}{\alpha_{\textnormal{s}}}
\newcommand{\gev}{\textnormal{GeV}}
\newcommand{\fm}{\textnormal{fm}}

\newcommand{\fig}{Fig.~}
\newcommand{\figs}{Figs.~}
\newcommand{\eq}{Eq.~}
\newcommand{\se}{Sec.~}
\newcommand{\eqs}{Eqs.~}

\title{Models for exclusive vector meson production in heavy-ion collisions}

\author{{\slshape T. Lappi$^{1,2}$, H. M\"antysaari$^1$}\\[1ex]
$^1$Department of Physics, 
 P.O. Box 35, 40014 University of Jyväskylä, Finland
\\
$^2$Helsinki Institute of Physics, P.O. Box 64, 00014 University of Helsinki,
Finland
}

\contribID{47}


\acronym{EDS'13} 

\maketitle

\begin{abstract}
We discuss coherent and incoherent photoproduction of $J/\Psi$ vector
mesons in high energy heavy ion collisions. In a dipole picture for the 
photon both can be naturally related to the dipole cross section that
is also probed in inclusive DIS. We compare results of a particular 
calculation to ALICE data.
\end{abstract}

\section{Introduction}

The dipole picture~\cite{Mueller:1989st,*Nikolaev:1990yw}
provides a very powerful tool to study QCD scattering at high energy or small~$x$. 
In this picture, the target is described by an energy dependent 
predominantly imaginary dipole-target scattering amplitude, commonly referred 
to as the ``dipole cross section'' $\sigmadip$.
The picture arises naturally in the Color Glass Condensate (see
e.g.~\cite{Gelis:2010nm,*Lappi:2010ek} for a review) description of high energy QCD, where $\sigmadip$
is a correlation
funtion of two fundamental representation Wilson lines, integrated over the transverse
plane.
Through this connection the  DIS $\sigmadip$ is related to 
 inclusive particle production, correlations and the thermalizing
matter in collisions of hadrons and nuclei at high energy (see 
e.g. \cite{Lappi:2012nh,*Lappi:2013zma}). 
In the dilute limit of small dipoles, on the other hand, it is proportional
 to the conventional integrated gluon distribution.

 We will here briefly describe, following \cite{Lappi:2010dd,Lappi:2013am},
 one particular application
of the dipole picture, 
namely on the calculation of vector meson production in ultraperipheral 
ion-ion collisions, which has been measured by the ALICE
collaboration~\cite{Abelev:2012ba,Abbas:2013oua}.
 The cross section is conventionally calculated by convoluting photon-nucleus 
scattering  with the nuclear photon flux~\cite{Bertulani:2005ru}.
In the dipole picture the photon-nucleus cross section is obtained by convoluting
the $\gamma \to q\bar{q}$ light cone wave function with the dipole cross section.
For vector meson production the dipole state must then be projected onto 
a phenomenological vector meson wave function, we refer the reader 
to the more complete description e.g. in Ref.~\cite{Lappi:2010dd}.

\section{Dipole cross sections in nucleons and nuclei}

Our baseline parametrization for the dipole cross section is the KT or 
IPsat model~\cite{Kowalski:2003hm,*Kowalski:2006hc},
where the dipole cross section is given by a DGLAP-evolved gluon distribution,
multiplied by a Gaussian impact parameter profile $T_p(\bt)$ and unitarized by exponentiation.
It can be very straightfowardly generalized to nuclei by  replacing the impact
parameter profile in the exponent   by a sum of impact parameter
profiles for $A$ nucleons at positions $\bt_i$, i.e. 
$ T_p(\bt)  \to \sum_{i=1}^A T_p(\bt-\bt_i)$. Note that 
that now the dipole-nucleus $S$-matrix is the
product of independent scatterings off the different nucleons. 
This interpretation 
enables one to directly generalize any $\sigmap$
to a nuclear one. To facilitate the analytical manipulations 
we in practice approximate the proton 
$\bt$ profile by a factorized one $\sigma_{\mathrm{dip}} = 2  T_p(\bt) \ampli(\rt)$;
this restriction is absent in
 the MC event generator SARTRE~\cite{Toll:2012mb}.

The nucleon coordinates $\bt_i$  must be averaged 
with  the standard  Woods-Saxon distribution. This average is here denoted by 
$\Aavg{\mathcal{O}(\{ \bt_i \})}$. 
With an explicit nucleon coordinate dependence 
one can calculate coherent and incoherent processes
consistently in the same parametrization. 
\begin{wrapfigure}{r}{0.5\textwidth}
\includegraphics[width=0.5\textwidth]{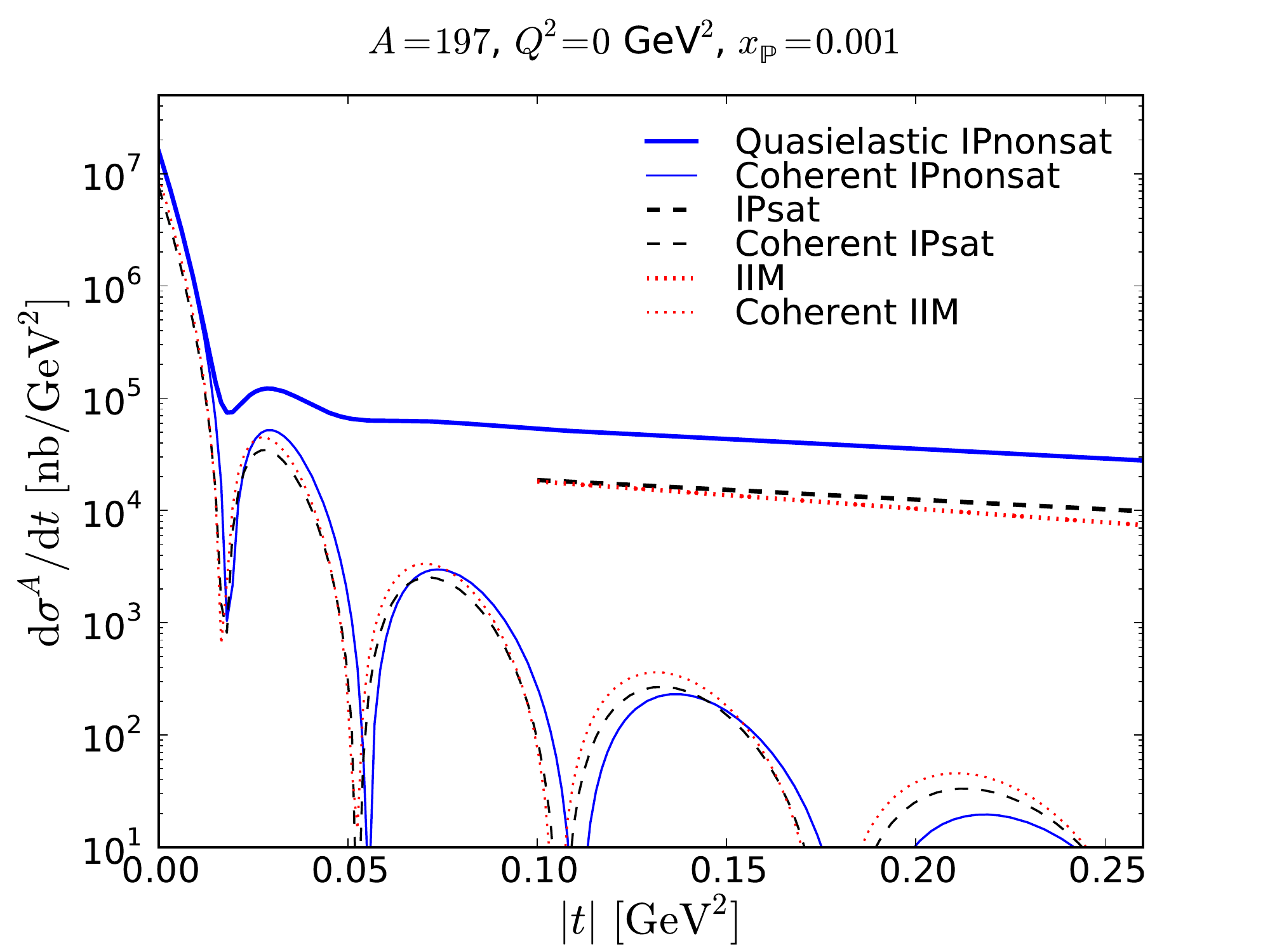}
\caption{$t$-dependence of the coherent and incoherent photoproduction cross sections.
}
\label{fig:tdep}
\end{wrapfigure}
The total quasielastic cross section 
is obtained by integrating over all momenta of the final state nucleons,
making the coordinates $\bt_i$ equal in the amplitude
and the complex conjugate, leading to 
$ \Aavg{|\A_{\gamma A \to J/\Psi A}|^2}$. For the coherent cross section, on the other hand,
we require the process to be fully elastic for the nucleus, i.e. 
$\bt_i$ must be the same in the initial and final states, and therefore
 independent in the amplitude and the complex conjugate, leading to
$ |\Aavg{\A_{\gamma A \to J/\Psi A}}|^2$. The incoherent cross section 
is the difference between the two, $ \sim \Aavg{|\A_{\gamma A \to J/\Psi A}|^2} -|\Aavg{\A_{\gamma A \to J/\Psi A}}|^2$,
and describes the fluctuations of 
the nucleon coordinates

\begin{figure}
\includegraphics[width=0.5\textwidth]{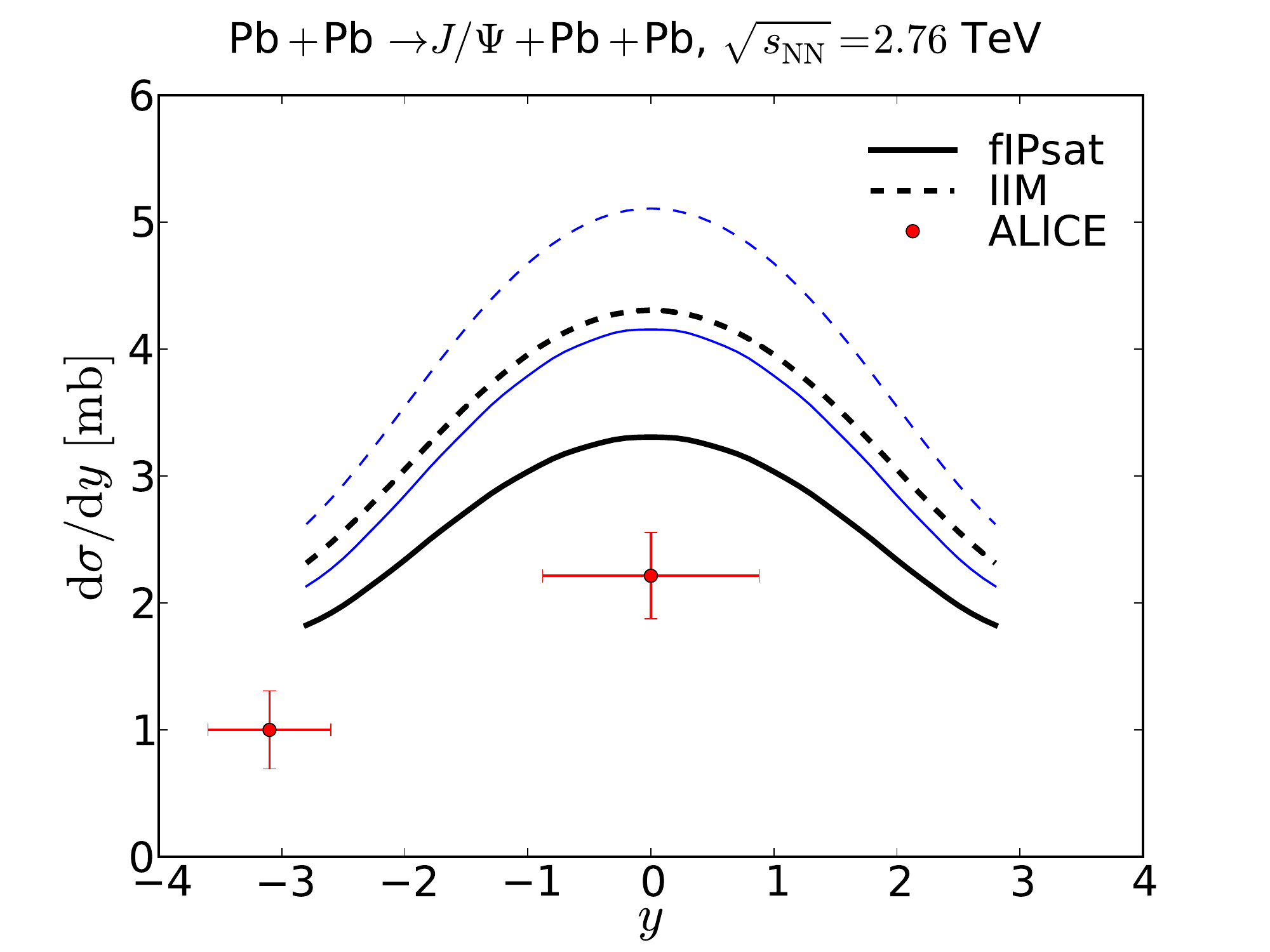}
\includegraphics[width=0.5\textwidth]{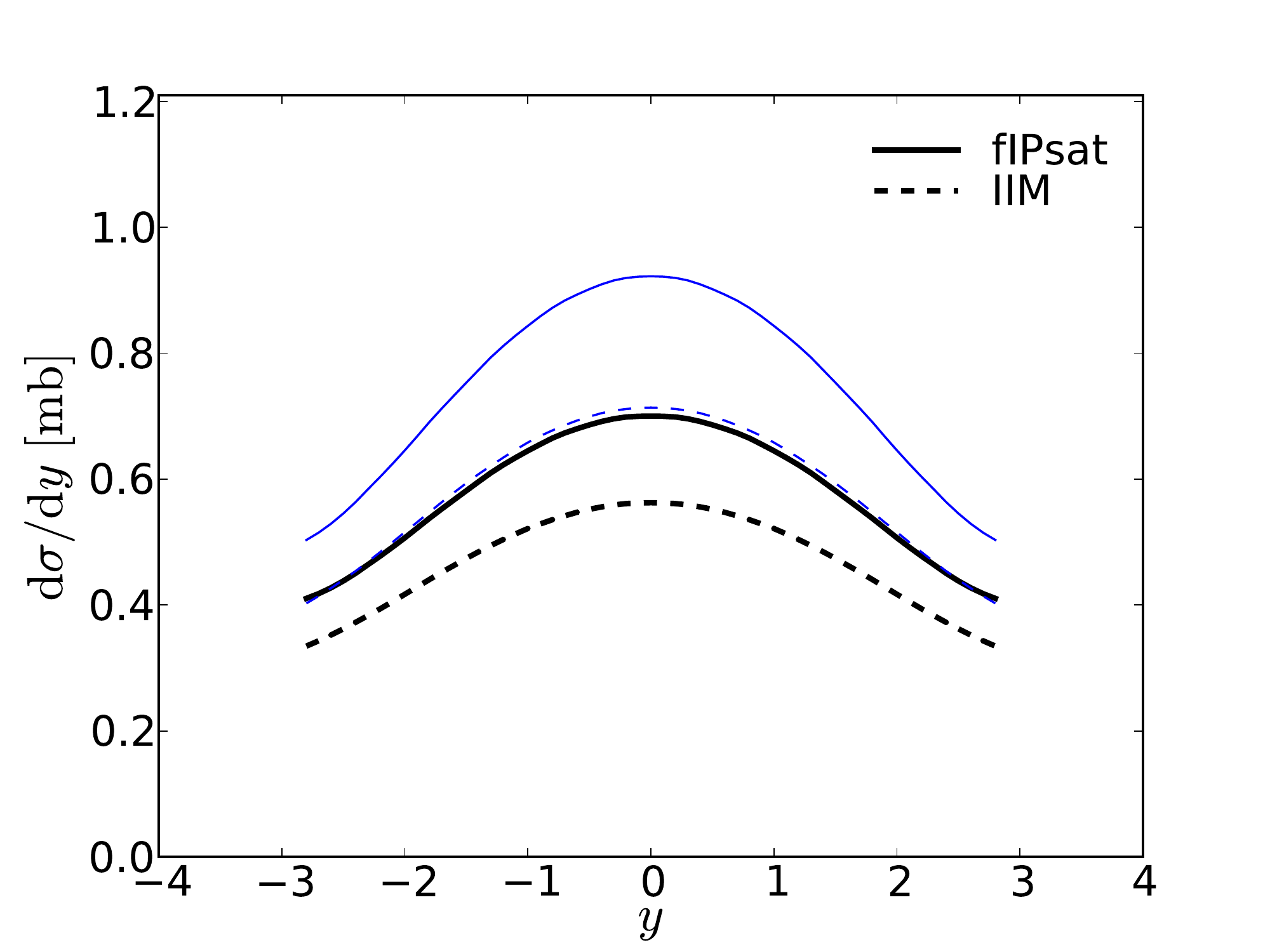}
\caption{Coherent (left) and incoherent (right) $J/\Psi$ production 
cross sections vs. rapidity for LHC kinematics, from \cite{Lappi:2013am}.
Different lines represent different dipole cross sections and vector meson wave functions.}
\label{fig:lm}
\end{figure}

Performing these averages leads to explicit expressions for the 
coherent and incoherent cross sections. We refer the reader to 
Eqs.~(12) and (13) in Ref.~\cite{Lappi:2013am} for the formulae
 (see also~\cite{Kopeliovich:2001xj}).
The incoherent cross section has a nice
interpretation as $A$ times the  $\gamma$ proton cross section
times a factor $\sim e^{- \sigma T_A}$
that enforces the requirement that the dipole must \emph{not}
scatter inelastically off the other $A-1$ nucleons.

\section{Results}

\begin{wrapfigure}{R}{0.45\textwidth}
\includegraphics[width=0.45\textwidth]{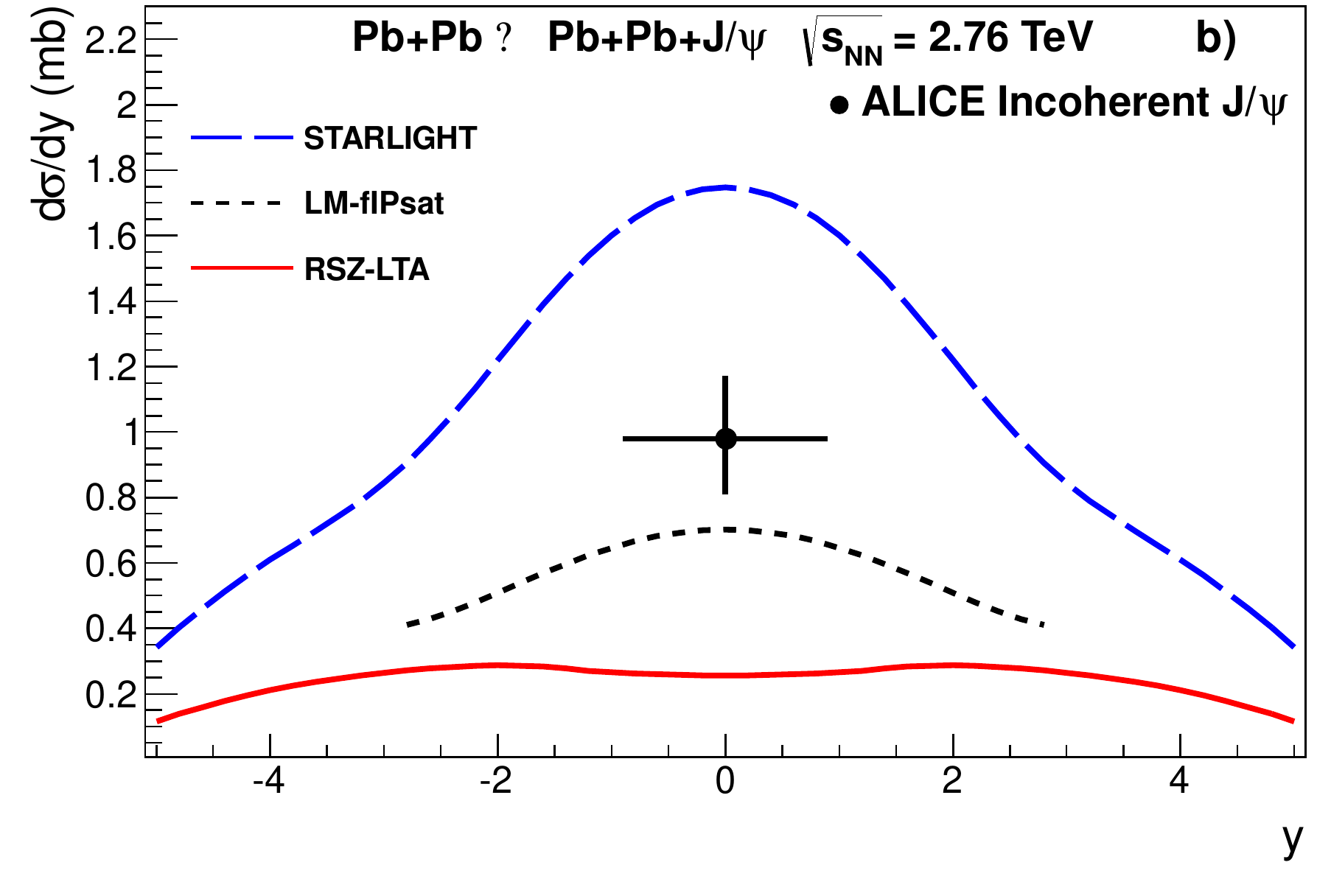}
\caption{ALICE incoherent cross section~\cite{Abbas:2013oua} compared to predictions.}\label{fig:aliceincoh}
\end{wrapfigure}
The $t$-dependence of the cross section 
is shown in \fig\ref{fig:tdep}. Note in particular the large 
suppression in the incoherent cross section compared $A \sigma_p $ 
(``IPnonsat''). Figure~\ref{fig:lm} shows the total cross section 
from~\cite{Lappi:2013am}. 
One sees a much stronger dependence on
the choice of the $J/\Psi$ wavefunction than for larger $Q^2$. The 
prediction overshoots the data, to a very large extent due to the
large skewness correction. 

Figure~\ref{fig:aliceincoh} shows a theory comparison of the 
\mbox{ALICE} incoherent cross section. The dipole model 
calculation~\cite{Lappi:2013am} is labeled ``LM'' in the figure.
The ``LTA''
calculation~\cite{Rebyakova:2011vf} differs mostly due to a different 
model of the nuclear breakup process. 
In the STARLIGHT event generator the incoherent cross section is
calculated by reducing $A\sigma^{\gamma p\to J/\Psi p}$  by a 
shadowing correction. Also in~\cite{Lappi:2013am} shadowing in the
coherent and incoherent cross sections are related to each other,
but by a more complicated functional form:
$\sim 1-\exp(-\sigma)$ for the coherent vs.  $\sim \sigma \exp(-\sigma)$ for the 
incoherent one, leading to a larger suppression for the latter. The 
data seems to favor the dipole picture here.

In conclusion, although one of the authors was told at this conference that 
the problem of $J/\Psi$ photoproduction off nuclei ``was solved 20 years ago,''
there are still
differences of a factor of $\sim 5$ between theoretical approaches, 
and the LHC data is an important constraint.

\paragraph{Acknowledgments}

 This work has been supported by the Academy of Finland, projects
133005, 267321, 273464 .
H.M. is supported by the Graduate School of Particle and Nuclear 
Physics.


\begin{footnotesize}

\bibliography{spires}
\bibliographystyle{h-physrev4mod2Mdense}

\end{footnotesize}
\end{document}